\documentclass[]{raa}                  
\usepackage{graphicx,times}             
\input{epsf.sty}                        
\input{psfig.sty}                       
\usepackage{mathrsfs}
\usepackage{amssymb}
\usepackage{txfonts}
\usepackage{threeparttable}
\begin{document}

   \title{A multiwavelength study of massive star-forming region IRAS 22506+5944
}
   \volnopage{Vol.0 (200x) No.0, 000--000}      
   \setcounter{page}{1}           

   \author{Yuan-Wei Wu \mailto{}
      \inst{1},
      Ye Xu \inst{1},
     Ji Yang \inst{1}
     \and
     Jing-Jing Li \inst{2}
      }

   \institute{$^{1}$Purple Mountain Observatory, Chinese Academy of Sciences,
             Nanjing 210008, China\\$^{2}$Shanghai Observatory, Chinese Academy of Sciences,
             Shanghai 200030, China
             \email{ywwu@pmo.ac.cn}
          }

   \date{Received~~2001 month day; accepted~~2001~~month day}
   \authorrunning{Wu et al.}            
   \titlerunning{Massive star-forming region IRAS 22506+5944}  

\abstract{We present a multi-line study of the massive star-forming
region IRAS 22506+5944. A new 6.7 GHz methanol maser was detected.
$^{12}$CO, $^{13}$CO, C$^{18}$O and HCO$^+$ J = 1-0 transition
observations reveal a star formation complex consisting mainly of
two cores. The dominant core has a mass of more than 200 M$_\odot$,
while another one only about 35 M$_\odot$. Both cores are obviously
at different evolutionary stages. A $^{12}$CO energetic bipolar
outflow was detected with an outflow mass of about 15 M$_\odot$.
   }

   \authorrunning{Yuan-Wei Wu \& Ye Xu  }            
   \titlerunning{Multi-line study of massive star-forming region IRAS 22506+5944 }  

   \maketitle

\section{Introduction}           
\label{sect:intro}

Massive stars play an important role in the evolution of the
interstellar medium (ISM) and galaxies; nevertheless their formation
process is still poorly understood  because of large distances, high
extinction, and short timescales of critical evolutionary phases. In
addition, massive stars do not form in isolation but often in
clusters and associations, which make the environment of massive
star formation regions more complex.

The 6.7 GHz transition of methanol has been found to be a
particularly useful signpost to trace massive star formation (Minier
et al.~\cite{mini03}, Xu et al.~\cite{yexu03}). On the other hand,
the maser phase encompasses the outflow phase (Xu et
al.~\cite{yexu06}), which give us another powerful tool to study the
dynamics of massive star formations.

IRAS 22506+5944, with an infrared luminosity of
1.5$\times$10$^4$L$_\odot$, belongs to the Cepheus molecular cloud
complex. Harju et al. (~\cite{harju93}) made a NH$_{3}$ map of this
region, and found a NH$_{3}$ core is coincident with the peak of the
IRAS source. Both H$_2$O maser (Wouterloot \& Walmsley
~\cite{wout86}) and SiO (Harju et al.~\cite{harju98}) have been
detected. Although searches for 6.7 GHz methanol maser (Szymczak et
al.~\cite{szym00}) show negative results, recently, we found a weak
6.7 GHz methanol maser in this region. Despite its high luminosity
and FIR color characteristics of the ultra-compact HII region, no
radio emission was detected (Molinari et al.~\cite{molinari98}). The
distances used in literatures for this source range from 5.0 kpc to
5.7 kpc. Here we use the value of 5.0 kpc.

In this paper, we present a multi-line study of this star-forming
region. In Sect. 2, we describe our observations. The results are
given in Sect.3. We give analysis and discussion in Sect.4, and
summarize in Sect.5.

\section{Observations}
\label{sect:observ}
\subsection{The Effelsberg 100 m Telescope}

Observations of the methanol (CH$_{3}$OH) maser were made using the
Effelsberg 100 m telescope in February 2006. The rest frequency
adopted for the 5$_1$-6$_0$ A$^+$ transition was 6668.519 MHz
(Breckenridge \& Kukolich ~\cite{brecken95}). The spectrometer was
configured to have a 10 MHz bandwidth with 4096 channels yielding a
spectral resolution of 0.11 km s$^{-1}$ and a velocity coverage of
450 km s$^{-1}$. The half-power beam width was $\sim$ 2$^\prime$ and
the telescope has an rms pointing error of 10$^\prime$$^\prime$. The
observations were made in position switched mode. The system
temperature was typically around 35 K during our observations. The
flux density scale was determined by observations of NGC7027 (Ott et
al.~\cite{Ott94}). The absolute calibration for flux density is
estimated to be accurate to $\sim$ 10\%. The integration time on
source was 10 minutes, with a rms noise level of $\sim$ 0.05 Jy in
the spectra.  The pointing position was R.A.(J2000)
22$^h$52$^m$36.9$^s$, DEC.(J2000) =
+60$^\circ$00$^\prime$48$^\prime$$^\prime$.

\subsection{The PMO 13.7 m Telescope at Delingha}

The $^{12}$CO, $^{13}$CO, C$^{18}$O and HCO$^+$ J = 1-0 maps were
observed with the PMO 13.7 m millimeter-wave telescope at Delingha,
China, during 2008 November. A cooled SIS receiver was employed, and
system temperatures was $\sim250$ K during the observations. Three
AOS (acousto-optical spectrometer) were used to measure the J = 1-0
transitions of $^{12}$CO, $^{13}$CO, C$^{18}$O and the FFTS (Fast
Fourier Transform Spectrometer) were used to measure the HCO$^+$ J =
1-0 lines. All the observations were performed in position switch
mode. The pointing and tracking accuracy was better than
10$^\prime$$^\prime$. The obtained spectra were calibrated in the
scale of antenna temperature T$^*_A$ during the observation,
corrected for atmospheric and ohmic loss by the standard chopper
wheel method. The grid spacings of the mapping observations were
30$^\prime$$^\prime$. Table\ref{Tab:obs para} summarizes the basic
information about our observations, including: the transitions, the
center rest frequencies $\nu_{rest}$, the half-power beam widths
(HPBWs), the bandwidths, the equivalent velocity resolutions
($\Delta\nu_{res}$), and the typical rms levels of measured spectra.
All of the spectral data were transformed from the T$^*_A$ to the
main beam brightness temperature T$^*_{MB}$ scale. The absolute
calibration for intensity was about 10\%.

The GILDAS software package (CLASS \& GREG) was used for the data
reduction.

\begin{table}[]
\caption[]{Observation Parameters}
\begin{center}\begin{tabular}{rrcccc}
 \hline
 \hline
Translation   & $\nu_{rest}$  &   HPBW  & Bandwidth  &    $\Delta\nu_{res}$      &  1$\sigma$ rsm$^a$ \\
  & (GHz)&($^\prime$$^\prime$)&(MHz)&(km s$^{-1}$)&(K) \\
 \hline
$^{12}$CO J = 1-0  &115.271204 &58 &145 &0.37 & 0.10 \\
$^{13}$CO J = 1-0  &110.201353 &61 &43 &0.11 & 0.10 \\
C$^{18}$O J = 1-0  &109.782182 &62 &43 &0.12 & 0.09 \\
HCO$^+$ J = 1-0     &89.188521  &75 &43 &0.16 & 0.10\\
 \hline
 \end{tabular}\end{center}
  \begin{tablenotes}
    \item [1]  \emph{$^a$} typical value in the scale of $T_{R}^{*}$.
 \end{tablenotes}
 \label{Tab:obs para}
\end{table}

\section{Results and Discussion}
\label{sect:results}
\subsection{spectra}
\subsubsection{6.7 GHz CH$_{3}$OH maser spectrum} 
\begin{figure}[h]
\noindent
\includegraphics[scale=0.5,angle=-90,bb=30 10 520 720]{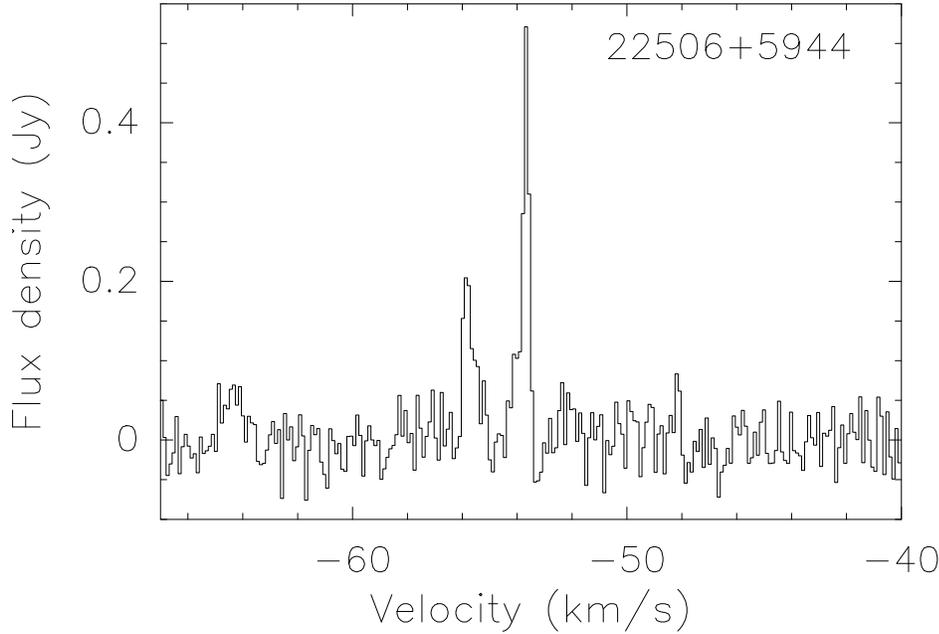}\caption{Spectrum of the
6.7-GHz CH$_{3}$OH maser. The spectral resolution is 0.11 km
s$^{-1}$.}\label{Fig:CH3OH spectra}
\end{figure}

The spectrum of the
CH$_{3}$OH maser detected in this region is shown in Fig.
\ref{Fig:CH3OH spectra}. There are two features that are separated
by about 2.2 km s$^{-1}$. The stronger feature is at the LSR (local
standard of rest) velocity of -53.7 km s$^{-1}$, with a flux density
of 0.52 Jy, while the other is only about 0.2 Jy. In order to get
high signal to noise spectra, we did not attempt to refine the
position and just integrated the time at the same position. Hence,
the actual position could be off by 1 arcminute.
\subsubsection{$^{12}$CO, $^{13}$CO, C$^{18}$O and HCO$^+$ spectra}
\begin{figure}[h]
\includegraphics[scale=0.8,bb=160 0 595 550,angle=-90]{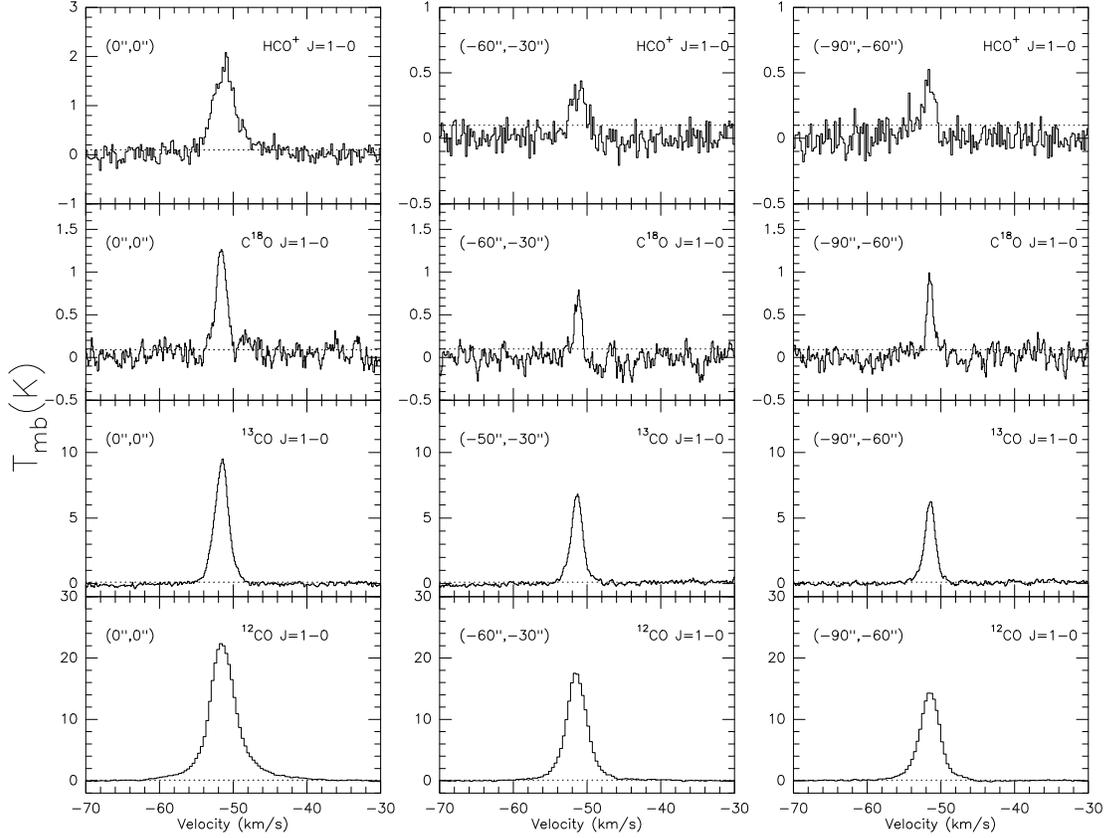}

\caption{\emph{Left panel:} spectra
 at the C$^{18}$O east peak. \emph{Middle panel:} spectra
at the conjunctive point of the two C$^{18}$O cores. \emph{Right
panel:} spectra at the C$^{18}$O west peak.  The horizontal dot line
is 1$\sigma$ level of each line (Table \ref{Tab:obs
para}).}\label{Fig:spectral}
\end{figure}

Spectra of $^{12}$CO, $^{13}$CO, C$^{18}$O and HCO$^+$ are presented
in Fig. \ref{Fig:spectral}. The spectra in left panel come from the
peak of core A (dominant core in Fig. \ref{Fig:mapping}). Both
$^{12}$CO and HCO$^+$ show remarkable broad line wings, with a FW
(full width) of 24 km s$^{-1}$ and 6 km s$^{-1}$ at 1$\sigma$ level,
respectively. Spectra in the right panel are correspondent to the
peak of core B, and spectra at the conjunctive point of the two
cores are given in middle panel. Details of the line in positions of
the peak, including the line central velocities, the fitted line
widths, the bright temperatures and integrated intensity were listed
in Table \ref{Tab:spectral}.


\begin{table}
 \caption{Result of molecular line measurements.}
 \begin{center}\begin{tabular}{ccccr}
 \hline
 \hline
Translation   & V$_{LSR}$  &   $\Delta\nu_{res}$  & T$^*_{MB}$  &    $\int$ T$^*_{MB}$ \emph{d}$\upsilon$  \\
 (GHz)        &(km s$^{-1}$)  &(km s$^{-1}$)  &(K)    &(K km s$^{-1}$) \\
 \hline
$^{12}$CO J = 1-0 \emph{$^a$}   &-51.4 &4.3 &22.3 &98.3  \\
$^{12}$CO J = 1-0 \emph{$^b$}   &-51.5 &3.3 &14.1 &48.9  \\
$^{13}$CO J = 1-0 \emph{$^a$}  &-51.4 &2.3 &9.5   &22.1  \\
$^{13}$CO J = 1-0 \emph{$^b$}  &-51.5 &1.8 &6.1   &11.4  \\
C$^{18}$O J = 1-0 \emph{$^a$}  &-51.6 &1.8 &1.2   &2.3   \\
C$^{18}$O J = 1-0 \emph{$^b$}  &-51.5 &1.1 &0.9   &1.0   \\
HCO$^+$ J = 1-0 \emph{$^a$}  &-51.1 &3.8   &2.1   &6.9  \\
HCO$^+$ J = 1-0 \emph{$^b$}  &-51.6 &2.0   &0.4   &0.9  \\
 \hline
 \end{tabular}
 \end{center}
 \begin{tablenotes}
    \item [1] \emph{a} and \emph{b} indicate Core A and
    Core B.
 \end{tablenotes}
 \label{Tab:spectral}
\end{table}
\subsection{Mapping}
\subsubsection{$^{13}$CO, C$^{18}$O and HCO$^+$ maps}
\label{subsubsect:clump conditions} 
\begin{figure}[h]
\centering
\includegraphics[bb=0 260 595 841,width=90mm,angle=90]{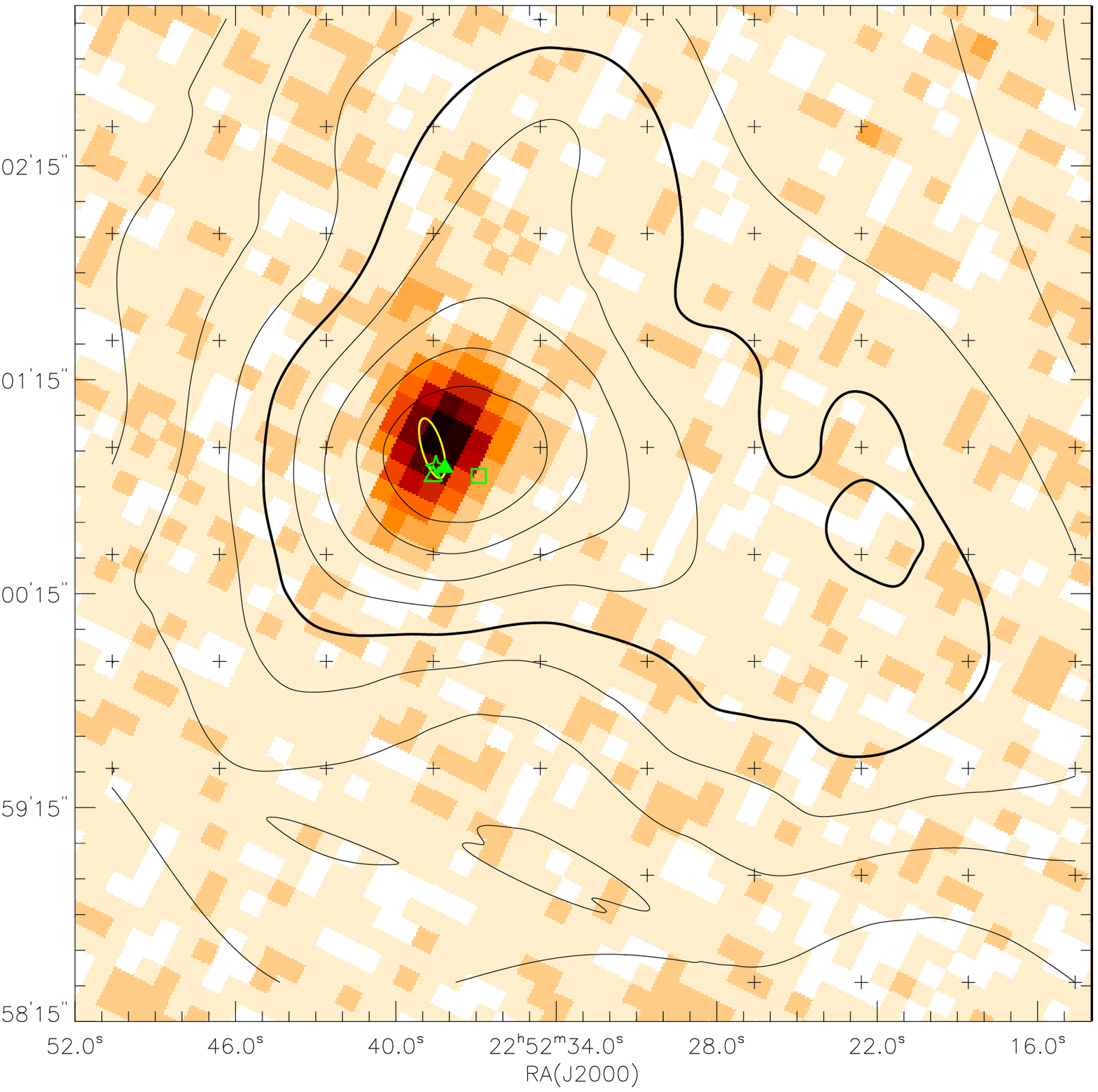}
\includegraphics[bb=0 260 595 841,width=90mm,angle=90]{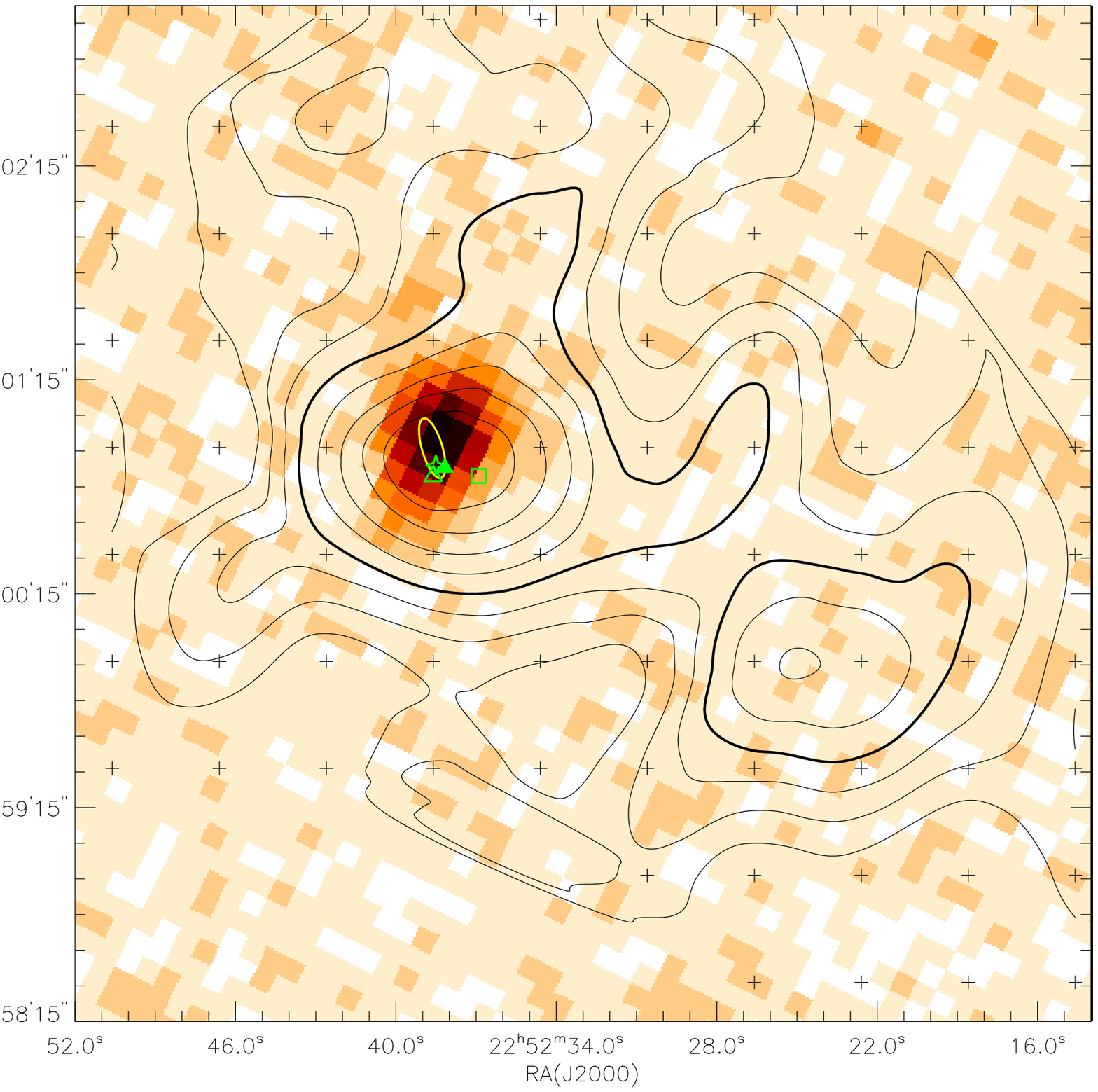}
\includegraphics[bb=0 260 595 841,width=90mm,angle=90]{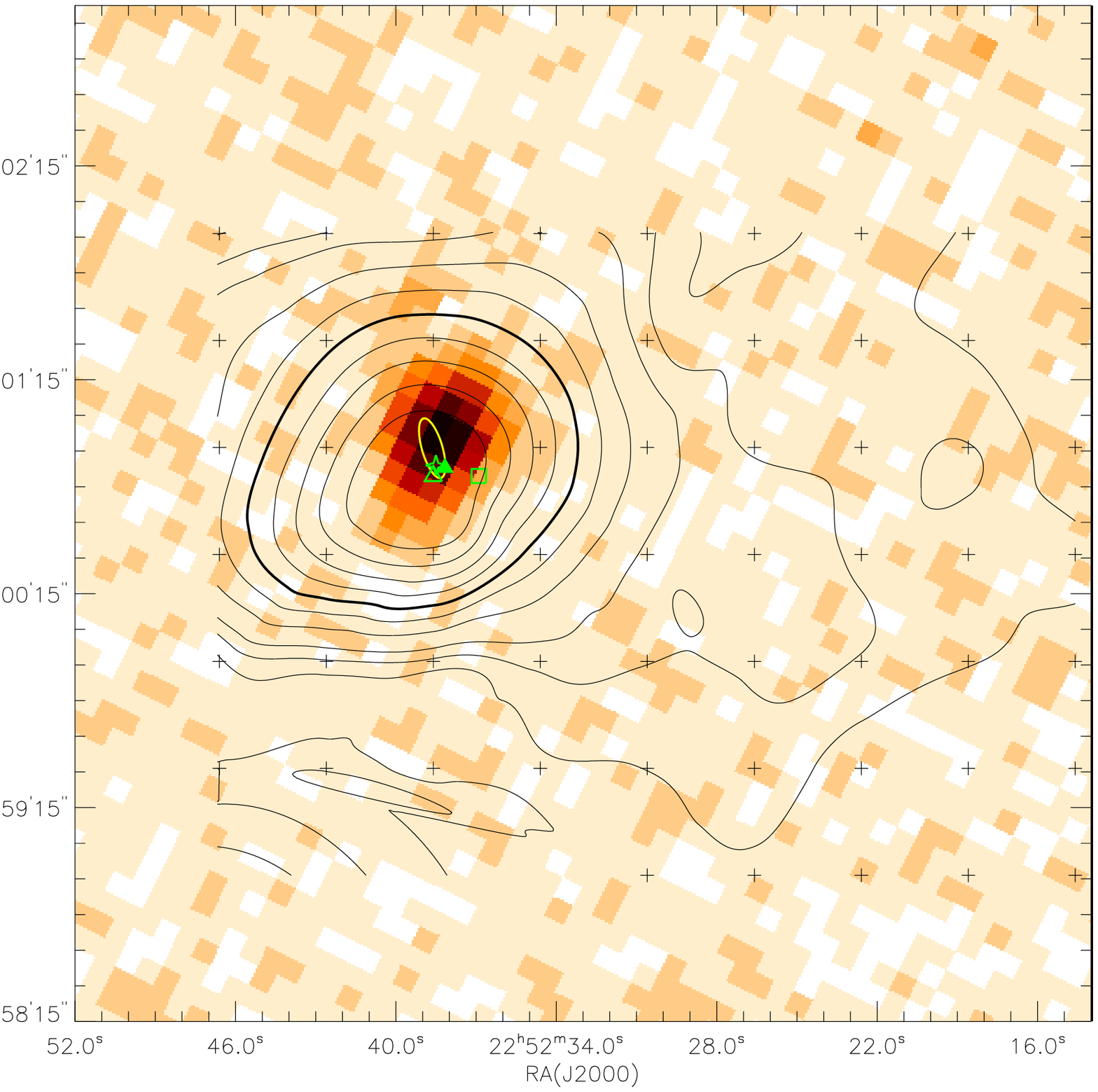}%
\includegraphics[bb=0 260 595 841,width=90mm,angle=90]{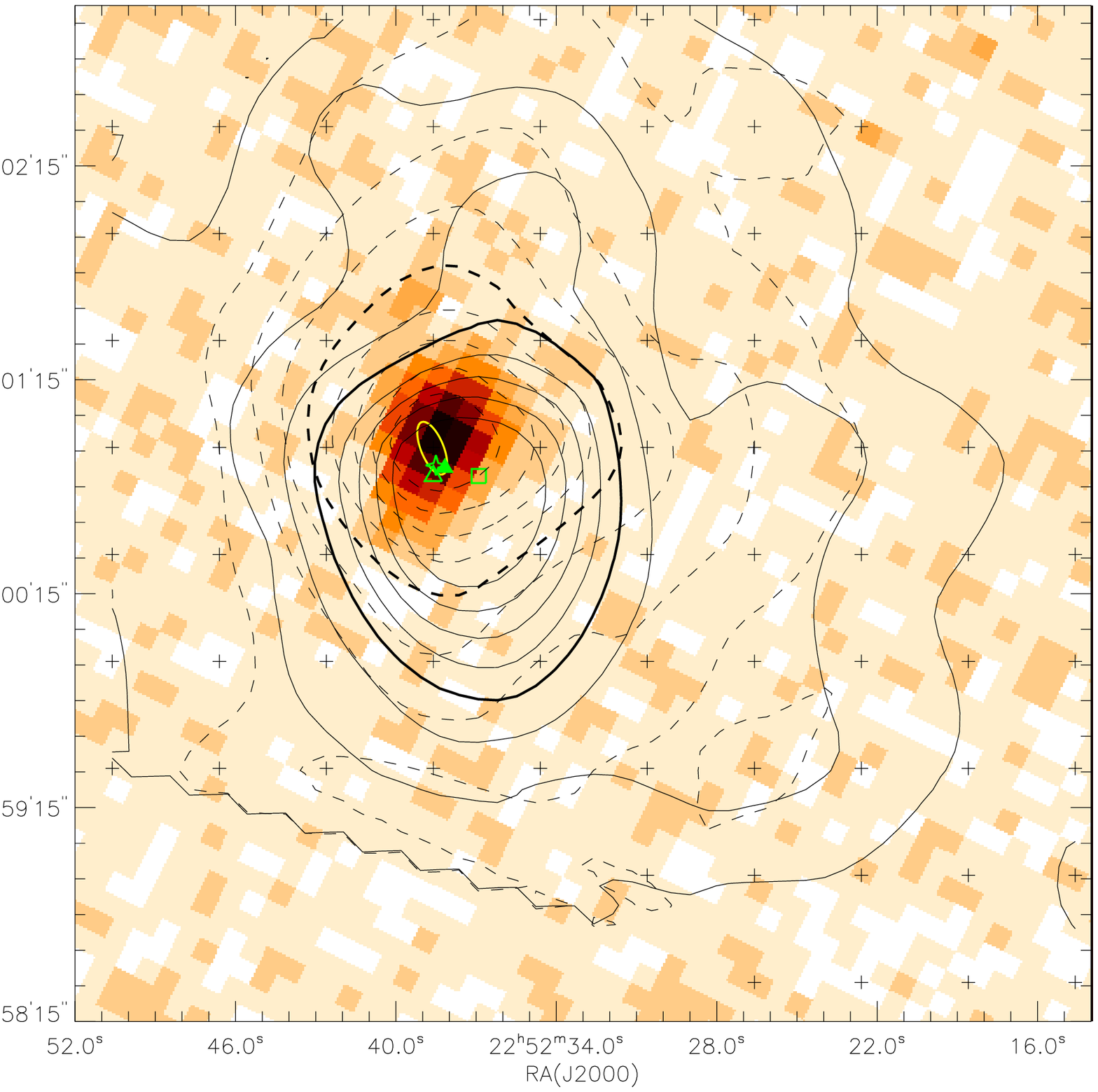}
\caption{\emph{upper left}: contour map of the total integrated
$^{13}$CO J = 1-0 line emission in the velocity range from -53.4 to
-49.5 km s$^{-1}$ overlaid on MSX E band 21 $\mu$m image.
\emph{upper right}: contour map of the total integrated C$^{18}$O J
= 1-0 line emission in the velocity range from -52.1 to -50.8 km
s$^{-1}$. \emph{lower left}: contour map of the total integrated
HCO$^+$ J = 1-0 line emission in the velocity range from -52.6 to
-49.5 km s$^{-1}$. \emph{lower right}: Contour map for the $^{12}$CO
J = 1-0 outflow. The blue wing (solid line) emission was integrated
over -60 to -54 km s$^{-1}$ and -49 to -42 km s$^{-1}$ for red wing
(dashed line),respectively. Contour levels of the plots are all from
20\% to 90\% by steps of 10\% of each peak emission with the
exception of HCO$^+$ J = 1-0 lines, whose contour levels are 10\%,
15\% and 20\% to 90\% by steps of 10\%. 50\% contour levels used to
determine core size are plotted with thicker lines. The small
crosses in the contour plots show the measured positions and the
ellipses mark IRAS error ellipse. The filled triangles denote 3 mm
continuum peak (Su et al 2004). H$_2$O (Wooterloot \& Walmsley
1986), SiO (Harju et al 1998) and CH$_{3}$OH masers are indicated
with open symbols of triangle, star and square, respectively.}
\label{Fig:mapping}
\end{figure}

Contour maps of the total
integrated $^{13}$CO J = 1-0, C$^{18}$O J = 1-0 and HCO$^+$ J = 1-0
line emissions were presented in Figure \ref{Fig:mapping}. We used
MSX E band (21$\mu$m) image as background images of the integrated
contours to compare the distributions between gas and dust. The
filled triangle denotes 3 millimeter continuum peak (Su et
al.~\cite{Su04}). H$_2$O (Wooterloot \& Walmsley ~\cite{wout86}),
SiO (Harju et al.~\cite{harju98}) and CH$_{3}$OH masers were
indicated with the open triangle, star and square, respectively. The
IRAS error ellipse is also marked. Contour levels are 20\% to 90\%
by steps of 10\% of the peak emission with the exception of HCO$^+$
J = 1-0 line, whose contour levels are 10\%, 15\% and 20\% to 90\%
by steps of 10\%.

\begin{figure}[h]
\centering
\includegraphics[width=150mm,angle=0]{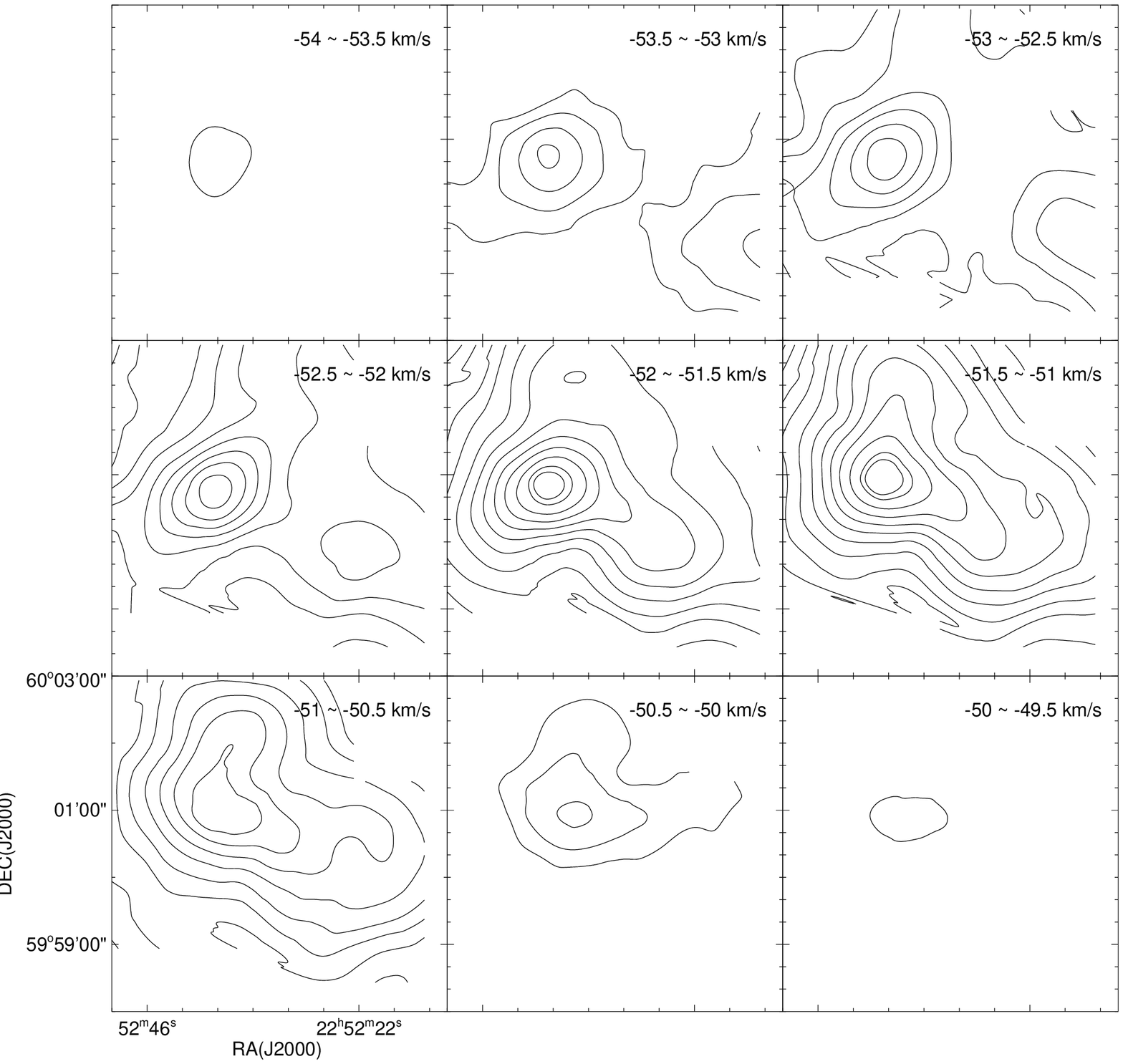}

\caption{Channel maps of $^{13}$CO J = 1-0 lines with contour levels
starting at 0.4 K km s$^{-1}$ and separated by 0.3 K km s$^{-1}$.}
\label{Fig:channelmap}
\end{figure}

From Figure \ref{Fig:mapping}, we see that both molecular line and
dust emission peak are roughly coincident with the IRAS source.
$^{13}$CO and HCO$^+$ are dominated with a single core. $^{13}$CO
shows a little elongation.  C$^{18}$O map clearly shows two cores
(Core A and Core B), indicating that such optical thin line traces
the inner part of a molecular cloud than other two lines. The center
of the Core A with an angular extent of (70$^\prime$$^\prime$,
60$^\prime$$^\prime$), coincides with IRAS 22506+5944 and MSX peak,
indicates that they may be the same source. The Core B has an offset
of (100$^\prime$$^\prime$, 60$^\prime$$^\prime$) at the north-west
of the Core A. The size of Core A is slightly larger than the Core B
(70$^\prime$$^\prime$, 60$^\prime$$^\prime$). In order to show the
kinematic relation of the two cores we also give the channel maps of
the $^{13}$CO J = 1-0 lines in Figure \ref{Fig:channelmap}.

We derive the physical parameters of the cores, assuming LTE (local
thermodynamic equilibrium) and with an abundance ratio
$[H_{2}]/[{}^{12}CO]$ = 10$^4$. Given a distance of the source to
galactic center, D$_{GC}$ $\sim $ 11.4 kpc ,we adopt an abundance
ratio $[{}^{12}CO]/[C^{18}O] \simeq 707 $ and
$[{}^{12}CO]/[{}^{13}CO] \simeq 93 $ estimated from the relationship
$[{}^{16}O]/[^{18}O] = (58.8\pm11.8)D_{GC}+(37.1\pm82.6)$ and
$[{}^{12}C]/[{}^{13}C] = (7.5\pm1.9)D_{GC}+(7.6\pm12.9)$ (Wilson \&
Rood ~\cite{wilson94}). Excitation temperature is calculated Using
equation \ref{eq:Tex}, assuming $^{12}$CO J = 1-0 lines are optical
thick:

\begin{equation}
T_{ex}^{*}=5.532\left\{
\ln\left[1+\frac{5.532}{\left(T_{R}^{*}\left(^{12}CO\right)+0.819\right)}\right]\right\}
^{-1} , \label{eq:Tex}
\end{equation}

$^{13}$CO and C$^{18}$O J = 1-0 line optical depths, $\tau $, are
estimated with formulas below:

\begin{equation}
\tau(^{13}CO)\thickapprox-\ln[1-\frac{T_{R}^{*}(^{13}CO)}{T_{R}^{*}({}^{12}CO)}]
\label{eq:tal13}
\end{equation}

\begin{equation}
\tau(C^{18}O)\thickapprox-\ln[1-\frac{T_{R}^{*}(C^{18}O)}{T_{R}^{*}({}^{12}CO)}]
\label{eq:tal18}
\end{equation}
$^{13}$CO and C$^{18}$O column densities are derived using equation
\ref{eq:N13CO} (Kawamura et al.~\cite{kawa98}) and equation
\ref{eq:NC18O} (Sato et al.~\cite{sato94}). $\tau$ and $\Delta\nu$
are optical depth and intrinsic line width:

\begin{equation}
N\left(^{13}CO\right)=2.42\times10^{14}\frac{T_{ex}\tau(^{13}CO)\Delta\nu(^{13}CO)}{1-\exp\left(-5.29/T_{ex}\right)}cm^{-2},
\label{eq:N13CO}
\end{equation}

\begin{equation}
N\left(C^{18}O\right)=2.24\times10^{14}\frac{T_{ex}\tau(C^{18}O)\Delta\nu(C^{18}O)}{1-\exp\left(-5.27/T_{ex}\right)}cm^{-2}.
\label{eq:NC18O}
\end{equation}

The nominal core size, \emph{l}, is determined by de-convolving the
telescope beam, using equation \ref{eq:size}:
\begin{equation}
l=D\left(A_{1/2}-\theta_{MB}^{2}\right)^{1/2} , \label{eq:size}
\end{equation}

where \emph{D} is the distance (5.0 kpc), \emph{A}$_{1/2}$ is the
area within the contour at the half- integrated intensity of the
peak and $\theta_{MB}$ is the main beam size (see Table \ref{Tab:obs
para}).

Core masses are computed with equation \ref{eq:mass}, where \emph{m}
is the mass of the hydrogen molecule, $\mu$ the ratio of total gas
mass to hydrogen mass, $\mu \approx$ 1.36 (Hildebrand
~\cite{hild83}), $N_{H_{2}}$ the column density of H$_2$ and
\emph{l} the de-convolved half power size defined above.

\begin{equation}
M_{LTE}=\mu mN_{H_{2}}l^{2}/4
 \label{eq:mass}
\end{equation}
The physical parameters derived are tabulated in Table \ref{Tab:phy
para}.

\begin{table}
 \caption{Physical parameters of Core A and Core B}
  \begin{center}\begin{tabular}{ccccrccccccccccc}
 \hline
 \hline
Name     & $\Delta\alphaup$  &    $\Delta\delta$ & \emph{l} &Tex & $\Delta \nuup$ \emph{$^a$} & N($^{13}$CO) &  N(C$^{18}$O)& N(H$_2$) \emph{$^b$} & M($_{LTE}$) \\
         & (arcsec)    &(arcsec) & (pc)& (K)&km s$^{-1}$ & (cm$^{-2})$ & (cm$^{-2}$)& (cm$^{-2}$) &(M$_\odot$)\\
 \hline
core A   & 90 & 70 & 0.9 & 26 & 1.5 &  2.7E+16 & 1.7E+15& 2.7E+22 & 228  \\
core B   & 85 & 60 & 0.5 & 18 & 1.0 & 10.7E+15 & 7.5E+14& 1.1E+22 & 35  \\

 \hline
 \end{tabular}\end{center}
 \begin{tablenotes}
    \item [1] \emph{a}: $\Delta \nuup$ has been corrected using
    $\frac{\Delta V_{line}}{\Delta
    V_{true}}=\sqrt{\frac{\ln\left[\tau/\ln\left(2/\left(1+e^{-\tau}\right)\right)\right]}{\ln2}}$
    ,considering line broadening due to optical depth.
    \item [2]    \emph{b}: H$_2$ column densities were derived using $^{13}$CO column
    densities, assuming $[H_{2}]/[{}^{13}CO]$ = 9.3$\times$10$^5$.

 \end{tablenotes}
 \label{Tab:phy para}
\end{table}
\subsubsection{Outflows}

Molecular outflows are an important signature of the earlier stage
in star formation. An outflow has been detected (Wu et
al.~\cite{wuyf05}) using the $^{12}$CO J = 2-1 line. A comparison of
two different transitions will be helpful to our better
understanding of the physical properties of outflows. In Fig.
\ref{Fig:mapping}, we present a similar work with the $^{12}$CO
(1-0) line.  The red and blue lobes are largely overlapped, while
the IRAS source is located at the center of the outflow, probably
the driving source of the outflow. Morphology of the $^{12}$CO J =
1-0 outflow is similar to that of the $^{12}$CO J = 2-1 outflow (Wu
et al.~\cite{wuyf05}), but the former extends a larger area than the
latter, spreading from Core A to Core B.

The outflow parameters, except for $^{12}$CO column density which is
derived from Snell et al. (~\cite{snell88}), are estimated with the
method of Beuther et al. (~\cite{beut02}). We assume that the gas is
in LTE and the line wings are optically-thin. Excitation temperature
and $[H_{2}]/[{}^{12}CO]$ abundance ratio adopted are the same as
Sect. \ref{subsubsect:clump conditions}. In order to better define
the kinematics of the high gas, we divided the wings into low
velocity and high velocity segments. The physical properties,
including velocity range, size, column density, mass, momentum and
kinetic energy are summarized in Table \ref{Tab:outflow para}.

Following the method of Beuther et al. (\cite{beut02}), we obtain
the characteristic time scale, \emph{t} $\approx 8.1 \times$ 10$^4$
yr, the mass loss rate, $\dot{M}_{out}\approx 1.8\times10^{-4}
$M$_\odot$ yr$^{-1}$ , the mechanical force, F$_m \approx
2.2\times10^{-3}$ M$_\odot$ km s$^{-1}$ yr$^{-1}$, and the
mechanical luminosity, L$_m \approx 2.2$ L$_\odot$. The mass and
kinetic energy of the outflow are significantly larger than typical
values from low-mass star forming regions (Bontemps et al. 1996).

\begin{table}
 \caption{Outflow properties}
  \begin{center}\begin{tabular}{cccccccccccccccc}
 \hline
 \hline
Compoent     & V$_{range}$  & Size\emph{$^a$} & N(H$_2$) & Mass & P & E$_k$\\
         & (km s$^{-1}$)     & (pc) & (cm$^{-2}$)& M$_\odot$ & (M$_\odot$ km s$^{-1}$) & (erg)\\
 \hline
red lobe(L)   &(-48.2 -44.0) & 0.8  &1.3E+20&5.3 &72 &9.6E+45 \\
red lobe(H)   &(-44.0 -38.0) & 0.8  &4.2E+19&2.1 &29 &3.8E+45\\
blue lobe(L)  &(-54.7 -59.0) & 1.0  &1.0E+20&6.6 &68 &7.1E+45\\
blue lobe(H)  &(-59.0 -62.0) & 1.0  &1.7E+19&1.0 &11 &1.1E+45\\
total         & ---          & ---- &---    &15.0&180&2.2E+46\\

 \hline
 \end{tabular}\end{center}
 \begin{tablenotes}
    \item [1]\emph{a}: size of lobes are computed using formula \ref{eq:size}.
 \end{tablenotes}
 \label{Tab:outflow para}
\end{table}

\subsection{evolutionary scenario}
HCO$^+$ usually traces the geometrically thick envelope of a core,
while C$^{18}$O is expected to trace the inner part of the core. The
C$^{18}$O map clearly shows two cores, Core A and Core B, which
likely consist of two star forming regions. Core A has obvious star
forming evidences, such as strong middle and far infrared emission,
masers and outflows, while Core B is only associated with some cold
molecular lines, indicating that core A and core B are in different
evolutionary stages, Core A at the phase of protostar core, while
Core B probably at the phase of pre-stellar core. The mass of Core A
is more than 200 $M_{\odot}$, while the IRAS source has a luminosity
of 1.5 $\times 10^{4}$ $L_{\odot}$. According to the relation
between mass and luminosity,
\begin{math} L \sim M^{3.5} \end{math}, the core mass is around one order of magnitude larger than
that of the IRAS source. This indicates there are other sources
within the core, which are not detected due to the resolution limit
of the employed telescope. With a rising steep spectrum, the IRAS
colors imply that the source is deeply embedded in a dense molecular
cloud. This IRAS source could be the exciting source of the H$_2$O,
SiO and CH$_{3}$OH masers. Core B has a mass of about 35
$M_{\odot}$, which might form several low or/and medium mass stars
in the future. The energetic outflow driven by IRAS 22506+5944
covers the whole region, including both Core A and Core B, and could
greatly affect its surrounding and accelerate Core B to form stars.
In summary, the whole region is a star formation complex, in which
stars at different evolutionary stages live in the same cluster and
interact with each other.

\section{Summary}
\label{sect:summary}

Our multi-line study reveals a star formation complex around IRAS
22506+5944, in which a weak 6.7 GHz CH$_{3}$OH maser was detected.
Multi-line Maps reveal a two-core structure: core A with mass of
$\sim$230 M$_\odot$ contains the IRAS source which is driving an
energetic bipolar outflow, while Core B, significantly smaller than
core A, has a large offset from the IRAS source. The two cores are
at different evolutionary stages. The energy released by the more
evolved core (Core A) is influencing the relatively less evolved
core (Core B) to accelerate its step to form stars.

\begin{acknowledgements}

We wish to thank all the staff at Qinghai Station of Purple Mountain
Observatory for their assistance with our observations. This work
was supported by the National Natural Science Foundation of China
(Grant Nos. 10673024, 10733030, 10703010 and 10621303) and National
Basic Research Program of China-973 Program 2007CB815403.
\end{acknowledgements}

\label{lastpage}

\end{document}